\documentclass{optica-article}

\journal{opticajournal} 

\articletype{Research Article}

\usepackage{lineno}
\usepackage[utf8]{inputenc}
\usepackage[T1]{fontenc}
\usepackage{amsmath}
\usepackage{graphicx}
\usepackage{xcolor}
\usepackage{float}

\usepackage{hyperref}
\hypersetup{
    colorlinks=true,
    citecolor=blue,
    linkcolor=blue,      
    urlcolor=blue,
}

\begin{document}

\title{Programmable and arbitrary-trajectory ultrafast flying focus pulses}

\author{M. V. Ambat,\authormark{1,2} 
J. L. Shaw\authormark{1} J.J. Pigeon,\authormark{1} K. G. 
Miller,\authormark{1} T. T. 
Simpson,\authormark{1} D. H. Froula,\authormark{1}and J. P. Palastro,\authormark{1,3} }

\address{\authormark{1}Laboratory for Laser Energetics, University of Rochester, Rochester, NY 14623, USA\\
\authormark{2}mvambat@ur.rochester.edu\\
\authormark{3}jpal@lle.rochester.edu}


\begin{abstract*} 
``Flying focus'' techniques produce laser pulses with dynamic focal points that travels distances much greater than a Rayleigh length. The implementation of these techniques in laser-based applications requires the design of optical configurations that can both extend the focal range and structure the radial group delay. This article describes a method for designing optical configurations that produce ultrashort flying focus pulses with arbitrary-trajectory focal points. The method is illustrated by several examples that employ an axiparabola for extending the focal range and either a reflective echelon or a deformable mirror-spatial light modulator pair for structuring the radial group delay. The latter configuration enables rapid exploration and optimization of flying foci, which could be ideal for experiments.

\end{abstract*}

\section{Introduction} \label{intro}
The intensity peak of a flying focus pulse can travel at any velocity, independent of the group velocity, over distances much longer than a Rayleigh range \cite{Sainte-Marie2017,Froula2018,Palastro2020,Jolly2020,Pigeon2023}. These properties offer a new approach to optimizing the wide range of laser-based applications that require velocity matching or extended interaction lengths. For instance, recent experiments have used a flying focus to create long, contiguous plasma channels \cite{Turnbull2018,Franke2019} and to synchronize the pump and probe pulses in soft x-ray lasers \cite{Kabacinski2023}. The potential uses of flying focus pulses  extend beyond these demonstrations to enhancing laser wakefield acceleration \cite{Palastro2020,Caizergues2020,Debus2019}, nonlinear Thomson scattering \cite{Ramsey2022}, or THz generation \cite{Simpson2023} and to facilitating observations of fundamental processes, such as radiation reaction \cite{Formanek2022} and Compton scattering \cite{Piazza2021}. The ultimate success of these applications relies on the design of practical, and preferably adaptive, optical configurations for preparing flying focus pulses.

The first experimental realization of a flying focus used a highly chromatic diffractive optic to focus a chirped laser pulse \cite{Froula2018}. The diffractive optic focuses each wavelength of the pulse to a different longitudinal location, while the chirp controls the arrival time of each wavelength at its focus. The resulting intensity peak traverses the focal range, i.e., the distance between the focal points of the minimum and maximum wavelengths, with a constant velocity that can be adjusted by changing the chirp. More complex spectral phases allow for more complex focal trajectories \cite{Sainte-Marie2017,PalastroIWAVs2018}. Despite its tunability, this ``chromatic flying focus'' has several limitations.  First, because the extended focal range is produced by a static diffractive optic, it  cannot be modified from shot to shot. Second and more importantly, the bandwidth of the pulse is spread across the focal region. This precludes the formation of an ultrashort (<100 fs) intensity peak, which is a requirement for many applications. 

\begin{figure}[ht!] 
\centering\includegraphics[width=\textwidth]{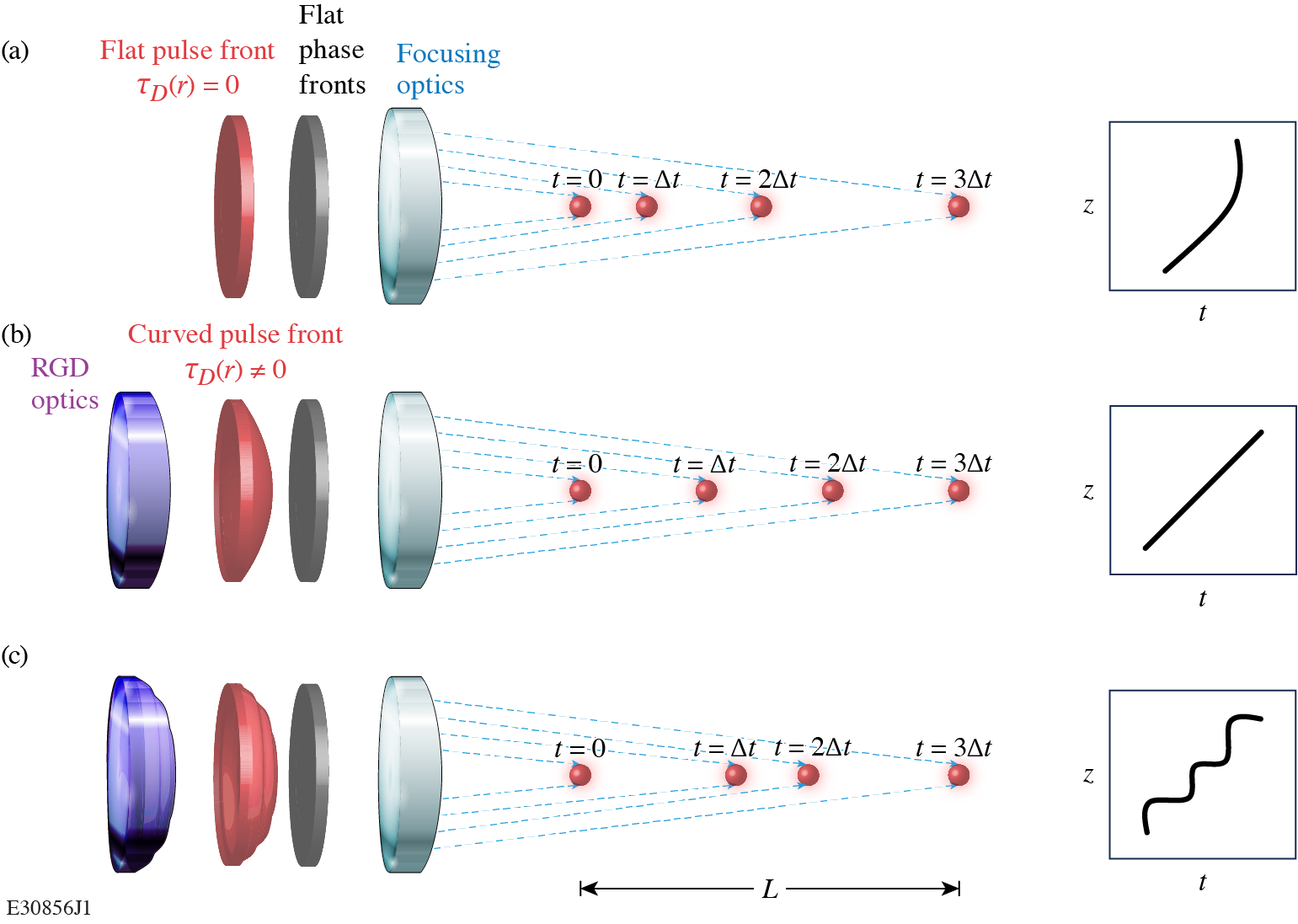} 
\caption{The effect of optics on the focal trajectory. (a) A laser pulse with a flat pulse front (red) and flat phase front (grey) is focused by an optic that extends the focal range $L$ (blue). The trajectory of the focus is completely determined by the focal geometry. (b) and (c) The pulse front, or radial group delay $\tau_D(r)$, is structured by a preliminary optic (purple). The structure of the pulse front can be used to create a constant-velocity focus (b), an oscillating focus (c), or otherwise dynamic trajectories.}
\label{fig:f1}
\end{figure}

The need for ultrashort intensity peaks has motivated the development of flying focus techniques that preserve the entire bandwidth of the laser pulse at every location within the focal range \cite{Palastro2020,Caizergues2020, Pigeon2023}. In contrast to the chromatic flying focus, which uses radial group delay to extend the focal range, these ``ultrafast flying focus'' schemes employ separate optics to independently extend the focal range and structure the radial group delay. As an example, a recent demonstration of an ultrafast, constant-velocity flying focus \cite{Pigeon2023} used the geometric aberration of an axiparabola  \cite{Smartsev2019,Oubrerie2022,Fan2023} to focus different annuli in the near field to different longitudinal locations in the far field and the radial group delay imparted by an echelon \cite{Palastro2020} to control the relative timing of the annuli. Despite the success of these experiments, the configuration relies on the use of a static echelon designed for a specific focal trajectory. An alternative configuration that replaces the echelon with adaptive optics, such as a deformable mirror-spatial light modulator pair \cite{Sun2015,Li&KawanakaDMSLM2020}, would allow for on-shot programmability of the radial group delay and, as a result, the focal trajectory. 

This work describes a method for designing optical configurations that produce ultrashort flying focus pulses with arbitrary focal trajectories at velocities close to the speed of light (Section II). The general method is independent of the optical configuration but is illustrated for specific examples of an axiparabola combined with either an echelon or a deformable mirror-spatial light modulator pair (Section III). The method is applied to create flying focus pulses exhibiting constant velocity, constant acceleration, and oscillating focal trajectories (Section IV). In each case, the intensity peak of the flying focus maintains an ultrashort duration as it traverses the extended focal range. The flexibility afforded by this method and the deformable mirror-spatial light modulator pair (DM-SLM) enable rapid and automated control over the focal trajectory, which can facilitate the use of the ultrafast flying focus in laser-based applications.

\section{The focal trajectory of an ultrafast flying focus} \label{derivation} 

Figure \ref{fig:f1} compares the trajectories of focal points produced by a focusing optic alone (a) and a focusing optic used in combination with optics that structure the radial group delay (b) and (c). In Fig. \ref{fig:f1}(a), a laser pulse with a flat phase front and a flat pulse front is incident at $z=0$ on a focusing optic with a surface defined by the sag function $s_f(r)$. The focusing optic extends the range of high intensity by using geometric aberration to focus different radial locations $r$ in the near field to different longitudinal locations in the far field $z = f(r)$. The resulting focal point travels a distance $L = \text{max}(f) - \text{min}(f)$ along a trajectory that is fully determined by the sag function. In Figs. \ref{fig:f1}(b) and (c), additional optics are used to structure the pulse front, or radial group delay $\tau_D(r)$, before focusing. Structuring the delay provides control over the trajectory of the focus and can produce a constant-velocity (b), oscillating (c), or otherwise dynamic focal point.

Each optical element in Fig. \ref{fig:f1} applies a spatio-spectral phase to the laser pulse. The phase imparted by the entire optical assembly $\phi(\omega,r)$ can be written as the sum of contributions from the focusing optic and the elements that structure the radial group delay (RGD). In the paraxial approximation (see Appendix A),
\begin{equation} \label{eq:phi}
\phi(\omega,r) =
 -\frac{2\omega}{c} s_f(r) + 
 \phi_D(\omega,r).
\end{equation}
The first term provides the initial phase front curvature required to focus each radius to the location $z = f(r)$. With $f(r)$ specified, the sag function $s_f(r)$ can be found by solving
\begin{equation} \label{eq:ds}
\frac{ds_f}{dr}= \frac{r}{2f(r)}.
\end{equation}
The second term in Eq. \eqref{eq:phi} modifies the relative timing of the near-field radii,
\begin{equation} \label{eq:tau}
\tau_D(r) = \frac{\partial\phi_D(\omega,r)}{\partial \omega}.
\end{equation}
To preserve the desired focusing, the elements that structure the RGD cannot significantly distort the phase fronts. The constraint $\partial_{r}\phi_D(\omega,r)|_{\omega=\omega_0} = 0$ ensures that $\phi_D$ only modifies the RGD and, equivalently, that the central frequency of the laser pulse $\omega_0$ focuses to the locations described by $f(r)$.  

For applications, one would like to specify a focal trajectory, i.e., the time-dependent velocity of the focus $\varv_f(t)$, and use this trajectory to determine the required $\tau_D(r)$. To calculate the required $\tau_D(r)$, first note that each near-field radius of the laser pulse can arrive at its focal location $z = f(r)$ at a different time. The focal time $t_f(r)$ for each radius has contributions from the structured RGD and the focal geometry:
\begin{equation} \label{eq:tf}
t_f(r) \approx \tau_D(r) + \frac{1}{c}\left[f(r) + \frac{r^2}{2f(r)} - 2s_f(r)\right].
\end{equation}
The variation in the focal time and location with radius results in a moving focal point with a velocity
\begin{equation} \label{eq:vf}
\tilde{\varv}_f(r) = \frac{df}{dr}\left(\frac{dt_f}{dr}\right)^{-1} \approx c\left[1 + \frac{r^2}{2f^2(r)} - c\left(\frac{df}{dr}\right)^{-1}\frac{d\tau_D(r)}{dr}\right].
\end{equation}
Equation \ref{eq:vf} demonstrates that the structured RGD can be used to control the trajectory of the focus independently of the focal geometry. If $\tau_D(r) = 0$, $\tilde{\varv}_f(r) = c[1+r^2/2f^2(r)]$, which is dictated soley by $f(r)$. Rearranging Eq. \eqref{eq:vf} provides a differential equation for the $\tau_D(r)$ needed to produce a specified trajectory $\varv_f(t)$:
\begin{equation} \label{eq:taudot}
c\frac{d\tau_D}{dr} = \left[1 - \frac{\varv_f\big(t_f(r)\big)}{c} + \frac{r^2}{2f^2(r)}\right]\frac{df}{dr},
\end{equation}
where $\varv_f\big(t_f(r)\big) = \tilde{\varv}_f(r)$ depends on $\tau_D$ through Eq. \eqref{eq:tf} and a one-to-one mapping between near-field radius and time has been assumed. The solutions to Eqs. \eqref{eq:ds} and \eqref{eq:taudot} form the basis for designing the optical elements necessary to create an ultrafast flying focus. 

In order to preserve the ultrashort duration of the intensity peak at every point within the focal range, the focal velocity must be close to the speed of light, $\varv_f(t) \approx c$. Even if a $\phi_D$ satisfies the constraint $\partial_{r}\phi_D|_{\omega=\omega_0} = 0$ and maintains the focal locations of the central frequency, it will modify the focal locations of every other frequency. This spreads the frequency content of the laser pulse across the focal region, which reduces the bandwidth available at each location and places a lower bound on the minimum duration. Noting that the transverse wavenumber is the radial derivative of the phase and using similar triangles, one can show that the RGD modifies the focal locations by a distance $\Delta f(\omega,r) \approx -cf^2(\partial_r\phi_D)/(r\omega)$. This longitudinal chromatism will have a negligible effect on the duration of the intensity peak when $\Delta f$ is much smaller than the focal range $L$, i.e., when
\begin{equation} \label{eq:durcond}
\frac{\Delta \omega}{\omega_0}\frac{f^2}{rL}\left| \frac{df}{dr}\left(1-\frac{\varv_f}{c} \right) \right| \ll 1,
\end{equation}
where $\Delta \omega$ is the bandwidth of the laser pulse and Eq. \eqref{eq:taudot} has been used with a simple form of $\phi_D(\omega,r) = (\omega-\omega_0)\tau_D(r)$.  


\section{Optical elements to create an ultrafast flying focus} \label{methods}

\subsection{Optics to extend the focal range} \label{focal range optics}

The optics that extend the focal range use geometric aberration to focus different radial locations $r$ in the near field to different longitudinal locations in the far field $z = f(r)$. In principle, this can be accomplished using refractive optics like lenses. However, for broadband, ultrashort pulses, the B-integral, group velocity dispersion, and higher-order dispersion of these optics can broaden or distort the temporal profile. In addition, the damage threshold of refractive optics typically prohibits their use as final focusing elements for high-intensity pulses. Thus, reflective optics are often preferable for extending the focal range of high-intensity, ultrashort flying focus pulses.

One such optic, the axiparabola \cite{Smartsev2019,Oubrerie2022}, produces an near-constant, on-axis intensity maximum over the entire focal range, making it ideal for many applications. The focal length as a function of near-field radius $f(r)$ is designed so that a flattop transverse intensity profile incident on the optic results in a uniform on-axis intensity maximum in the far field. Specifically, 
\begin{align} 
f(r) &= f_0 + L\left(\frac{r}{R}\right)^2, \label{eq:axipf} \\ 
s_f(r) &= \frac{R^2}{4L}\ln{\left[1 + \frac{L}{f_0}\left(\frac{r}{R}\right)^2\right]}, \label{eq:axips}
\end{align}
where $f_0$ is the nominal focal length, $R$ is the maximum radius of the axiparabola, and $L$ determines the length of the focal range. Expanding Eq. \eqref{eq:axips} in powers of $q\equiv L/f_0$ shows that the axiparabola is primarily a parabolic mirror $\mathcal{O}(q^0)$ with spherical aberration $\mathcal{O}(q^1)$. For $L>0$ ($<0$), rays incident at larger radii are focused farther from (closer to) the optic than rays incident at smaller radii. With this choice of $f(r)$, Eq. \eqref{eq:durcond} simplifies to $2(\Delta \omega/\omega_0)(f_0/R)^2|1-\varv_f/c| \ll 1$, which is independent of $L$.

\begin{figure}[ht!]
\centering\includegraphics[width=\textwidth]{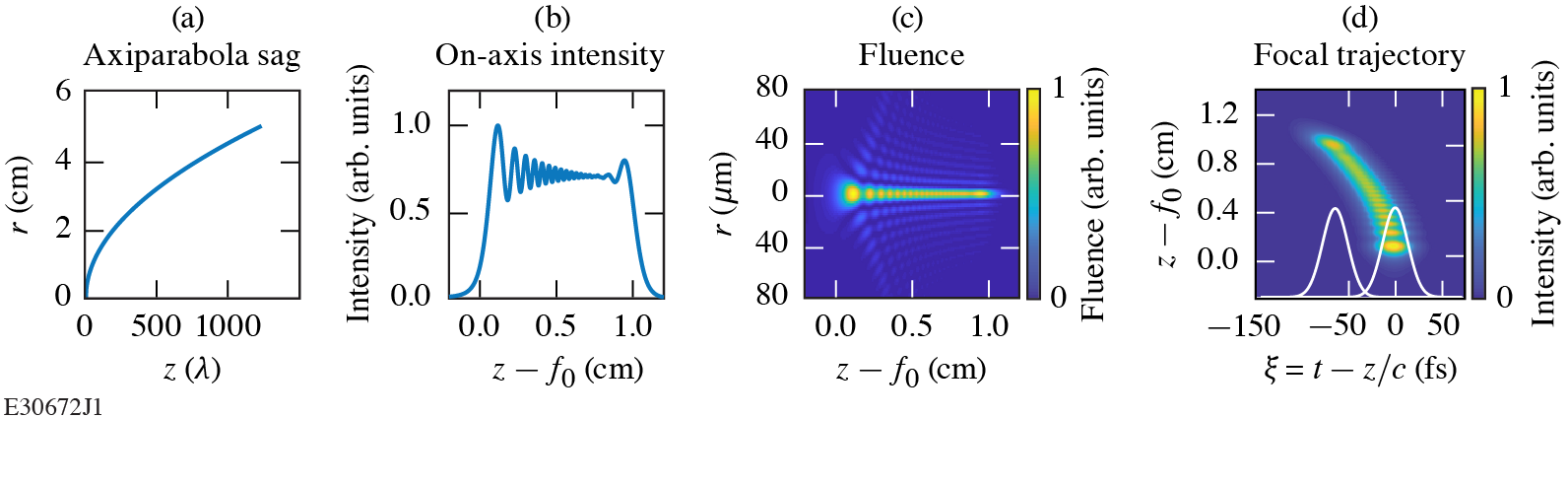}
\caption{The focal properties of an axiparabola alone. (a) The sag function of an axiparabola with $f_0=$ 50 cm, $R=$ 5 cm, and $L=$ 1 cm. The axiparabola focuses a laser pulse with a central wavelength of $\lambda_0 = 920$ nm and a 27 fs FWHM duration. (b) The maximum on-axis intensity of the pulse as a function of distance from the nominal focal point $z=f_0$. (c) The fluence profile. (d) The focal trajectory as a function of propagation distance and moving frame coordinate $\xi = t-z/c$. The peak intensity travels at a superluminal velocity and accelerates. The white lineouts show the temporal profile of the pulse at the beginning (right) and end (left) of the focal region.}
\label{fig:f2}
\end{figure}

Figure \ref{fig:f2} displays the results of propagation simulations (see Appendix B) for a laser pulse focused by an axiparabola with $f_0=$ 50 cm, $R=$ 5 cm, and $L=$ 1 cm. The laser pulse had a central wavelength $\lambda_0 = 2\pi c/\omega_0=$ 920 nm and $\Delta \lambda$ = 78 nm of bandwidth in a Gaussian power spectrum, corresponding to a 27 fs full-width at half-maximum (FWHM) duration. The transverse profile was initialized as a flattop with a 5 cm radius that filled the aperture of the axiparabola. The maximum on-axis intensity is nearly uniform over the entire focal range $L$, which is $\sim$$340\times$ longer than the Rayleigh range of the full-aperture focal spot $Z_R = \lambda_0f_0^2/\pi R^2$ [Fig. \ref{fig:f2}(b)]. The modulations in the on-axis intensity result from diffraction of the spherically aberrated phase fronts (see Appendix C). The near-uniform on-axis intensity comes at the cost of a spot size $w$ that narrows over the focal range [Fig. \ref{fig:f2}(c)]. More specifically, the effective $f/\#$ at the beginning of the focal range is larger than that at the end, such that within the focal region 
\begin{equation} \label{eq:wz}
w(z) \approx \frac{\lambda_0f_0}{\pi R}\left |\frac{L}{z-f_0}\right|^{1/2}.
\end{equation}
The ring-like structures visible in the fluence [Fig. \ref{fig:f2}(c)] are the natural diffraction pattern created by the axiparabola.

Figure \ref{fig:f2}(d) illustrates the focal trajectory produced by the axiparabola. Here, the on-axis intensity is plotted as a function of propagation distance $z-f_0$ and the moving frame coordinate $\xi = t-z/c$. In these coordinates, a vertical line indicates a signal travelling at the vacuum speed of light. The intensity peak accelerates from its initial focal point at $z-f_0 = 0$ and $\xi = 0$ to its final focal point at $z-f_0 = L$ and $\xi \approx -75$ fs, following a trajectory consistent with $\tilde{\varv}_f(r) = c[1+r^2/2f^2(r)]$. The pulse maintains its ultrashort duration over the entire focal range as shown by the white lineouts taken at the start (right) and end (left) of the focal region. 

\subsection{Optics to structure the radial group delay} \label{RGD optics}
The trajectory of the focus can be programmed by structuring the radial group delay of the laser pulse. Ideal, achromatic focusing optics impart the exact amount of RGD needed to ensure that all frequency components within a pulse arrive at their focus at the same time. More generally, optics can impart unwanted RGD, resulting in asynchronous focusing and a reduction in the maximum focused intensity. For instance, with refractive optics, the combination of group velocity dispersion and the radially dependent thickness of the optic produce unfavorable RGD \cite{Bor1988}. Below, optical elements are discussed that can impart favorable RGD, thereby enabling control over the trajectory of the focal point and the peak laser intensity. 

The recently proposed and demonstrated radial echelon provides a reflective approach to structuring the radial group delay \cite{Palastro2020,Pigeon2023}. The mirrored surface of the echelon consists of concentric rings with variable widths determined by the desired RGD and depths $d$ equal to a half-integer multiple of the central wavelength $d = (\ell/2)\lambda_0 = \pi\ell/\omega_0$, where $\ell$ is a positive integer. For a given $\tau_D(r)$ and $\ell = 1$, the phase imparted by the echelon is given by
\begin{equation} \label{eq:echphase}
\phi^{\text{ech}}_{D}(\omega,r) = -\frac{2\omega}{c} \left\{ \frac{1}{4} \lambda_0 \left[ \text{ceil} \left(\frac{c\tau_{D}(r)}{\lambda_0} \right) + \text{floor} \left( \frac{c\tau_{D}(r)}{\lambda_0} \right) \right] \right\}.
\end{equation}
By discretizing the continuous delay $c\tau_D(r)$ in steps of the central wavelength, the echelon satisfies the constraint $\partial_{r}\phi^{\text{ech}}_D(\omega,r)|_{\omega=\omega_0} = 0$ and thus does not affect the focusing of the frequency component $\omega_0$. Said differently, the phase fronts of the central wavelength maintain their transverse coherence upon reflection from the echelon. For any other wavelength, the echelon introduces a shear in the phase front between each ring. This shear smooths out as higher-spatial orders diffract, leaving the desired radial group delay. The widths of the echelon rings can also lead to diffractive losses. These losses are negligible when $\Delta R \gg \lambda_0f_0/2R$, which is easily satisfied for a large range of designs. Importantly, for $\varv_f(t) \approx c$, the combined axiparabola-echelon system preserves an ultrashort pulse duration. 

Despite its advantage as a reflective optic with a higher damage threshold, each echelon is a static optical element that can only impart a single, pre-designed RGD. Adaptive optics, such as deformable mirrors and spatial light modulators, offer dynamic programmability of the radial group delay and, as a result, the focal trajectory. A deformable mirror (DM) consists of pistons or piezoelectric segments that shape a flexible, reflective membrane \cite{Nemoto1997, Liu2013}. A DM can be programmed to apply the continuous phase 
\begin{equation} \label{eq:DMphase}
\Phi_{\text{dm}}(\omega,r) = -\frac{2\omega}{c}s_{\text{dm}}(r) = \omega \tau_D(r),
\end{equation}
where $s_{\text{dm}}(r) = -c\tau_D(r)/2$ is the sag function of the membrane. However, the phase $\Phi_{\text{dm}}(\omega,r)$ does not satisfy the constraint $\partial_r\Phi_{\text{dm}}(\omega,r)|_{\omega=\omega_0} = 0$. Thus a second optical element must be introduced to eliminate the phase distortion at the central frequency.

\begin{figure}[ht!]
\centering\includegraphics[width=\textwidth]{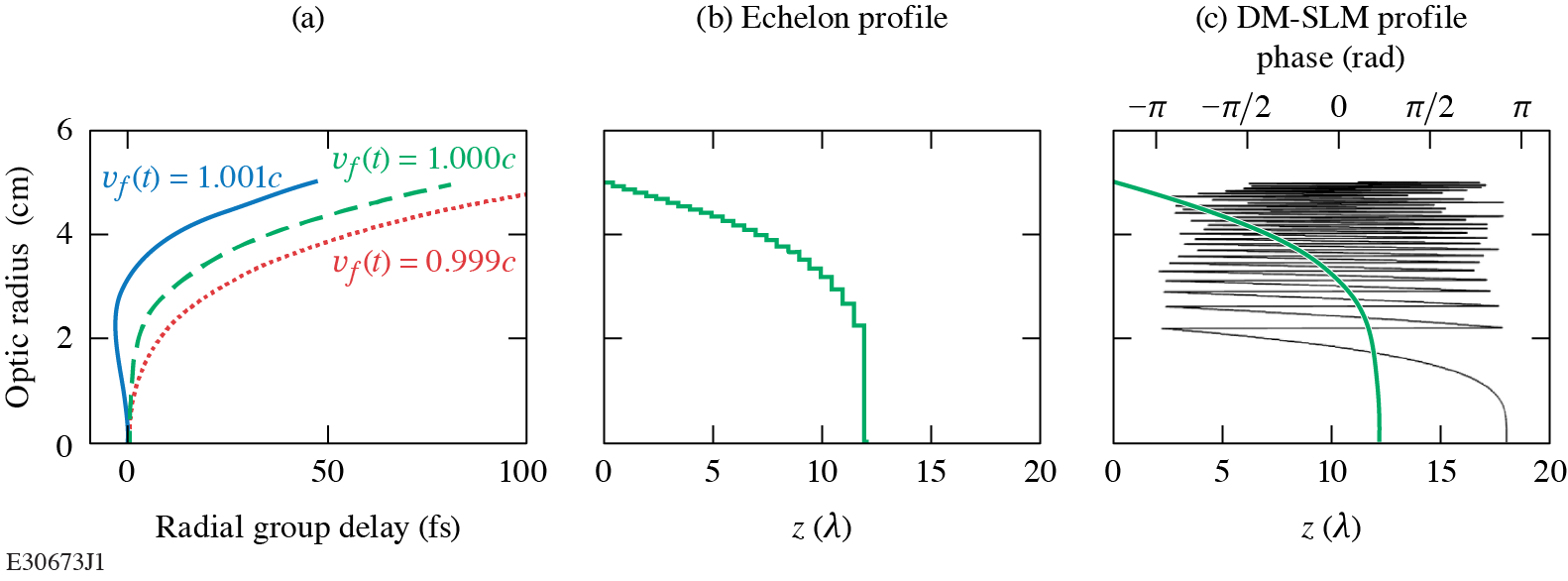}
\caption{(a) The radial group delays, i.e., the $\tau_D(r)$, required to produce constant-velocity focal trajectories with $\varv_f = 1.001c$, (blue, solid), $\varv_f = c$ (green, dashed), and $\varv_f = 0.999c$ (red, dotted) with the axiparabola described in Fig. \ref{fig:f2}.  (b) The echelon profile for $\varv_f = c$. (c) The deformable mirror sag function (green) and spatial light modulator phase (black) for $\varv_f = c$.}
\label{fig:f3}
\end{figure}

A spatial light modulator (SLM) can partially correct the phase front distortion at the central frequency \cite{Li&KawanakaDMSLM2020}.  An SLM consists of a pixelated, two-dimensional array of liquid crystals that possess electrical and optical anisotropy. The voltage delivered to each pixel can be adjusted to change the optical path length of an incident laser pulse as a function of transverse location \cite{Weiner2000, Bahk2014}. By appropriately programming the SLM voltages, the phase front of the central frequency can be flattened to an extent allowed by the discreteness of the pixels. Specifically, for the DM phase in Eq. \eqref{eq:DMphase},
\begin{equation} \label{eq:SLMphase}
\Phi_{\text{slm}}(\omega,r) = - \frac{\omega}{c} \lambda_0\text{mod} \left[ \frac{c\tau_{D} (r_p)}{\lambda_0}, 1 \right],
\end{equation}
where $r_p = \tfrac{1}{2}[\text{floor}(\tfrac{r}{p}) + \text{ceil}(\tfrac{r}{p})]p$ and $p$ is the SLM pixel size. The total phase of the DM-SLM pair is then 
\begin{equation} \label{eq:SLMDMphase}
\phi_D^{\text{dm-slm}}(\omega,r) = \Phi_{\text{dm}}(\omega,r) + \Phi_{\text{slm}}(\omega,r). 
\end{equation}
In the limit of infinitesimal pixels, $p\rightarrow0$ and $\phi_D^{\text{dm-slm}}(\omega,r)\rightarrow \phi_D^{\text{ech}}(\omega,r)$. Note that Eq. \eqref{eq:SLMphase} was discretized into radial zones; for Cartesian zones, one can instead use $\tau_{D}(x_p, y_p)$.

\begin{figure}[H]
\centering\includegraphics[width=\textwidth]{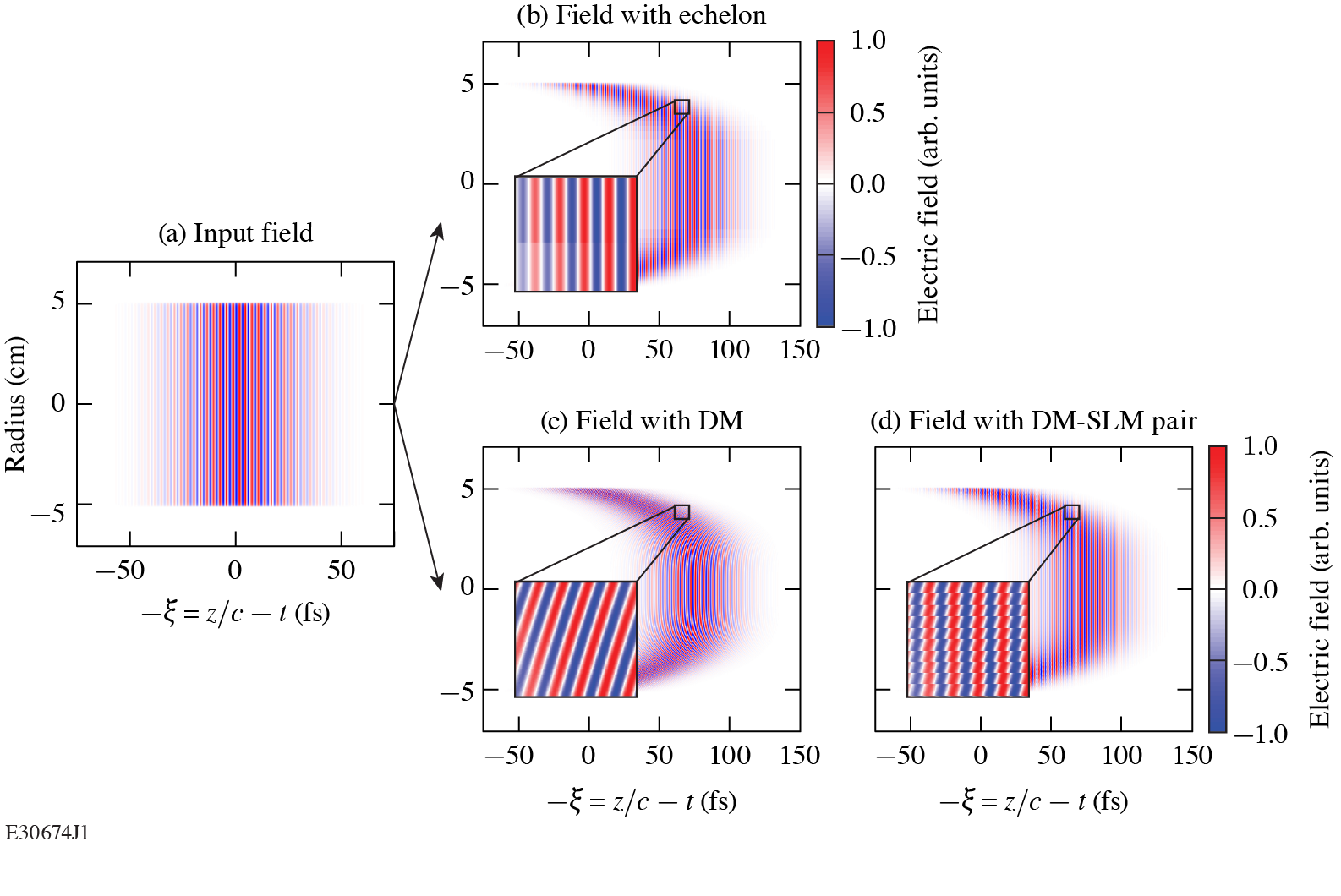}
\caption{Modification to the electric field in the near field for $\varv_f = c$. (a) The input field has flat phase fronts, a flat pulse front, and an ultrashort duration ($\sim$ 27 fs).  (b) The echelon imparts the desired radial group delay to the pulse while maintaining the flat phase fronts. (c) The DM imparts the desired radial group delay to the pulse. However, as shown in the inset, the phase fronts are now curved with respect to the propagation direction. (d) The SLM corrects the undesired phase front curvature. The inset shows that the phase fronts are now globally flat, but retain a residual tilt within each pixel. Each inset is a 500 $\mu$m $\times$ 15 fs window, and the SLM had a $p =$ 50 $\mu$m pixel size. The pulse propagates from left to right.}
\label{fig:f4}
\end{figure}

Figures \ref{fig:f3} and \ref{fig:f4} illustrate how these optics modify the electric field profile of a laser pulse in the near field to produce a constant-velocity focus. Figure \ref{fig:f3}(a) shows the $\tau_D(r)$ required for subluminal ($\varv_f <c$), luminal ($\varv_f = c$), and superluminal ($\varv_f>c$) focal velocities when using the axiparabola described in Fig. \ref{fig:f2}. Because the axiparabola naturally produces a superluminal and accelerating focus, the subluminal (superluminal) velocity requires a larger (smaller) delay than the luminal velocity at larger radii. The echelon and DM-SLM designs for $\varv_f = c$ are displayed in Figs. \ref{fig:f3}(b) and (c). In this configuration, the incident laser pulse propagates from right to left, so that the center of the pulse encounters the optics first. Figure \ref{fig:f4} shows the effect that each optic has on the electric field profile. After the echelon [Fig. \ref{fig:f4}(b)], the field has flat phase fronts and a radially dependent delay consistent with $\tau_D(r)$. After the DM [Fig. \ref{fig:f4}(c)], the field has the correct delay, but also has curved phase fronts. The SLM undoes this curvature [Fig. \ref{fig:f4}(d)]. The combined DM-SLM system reproduces the field profile created by the echelon to within the resolution limits of the SLM.

\begin{figure}[H]
\centering\includegraphics[width=\textwidth]{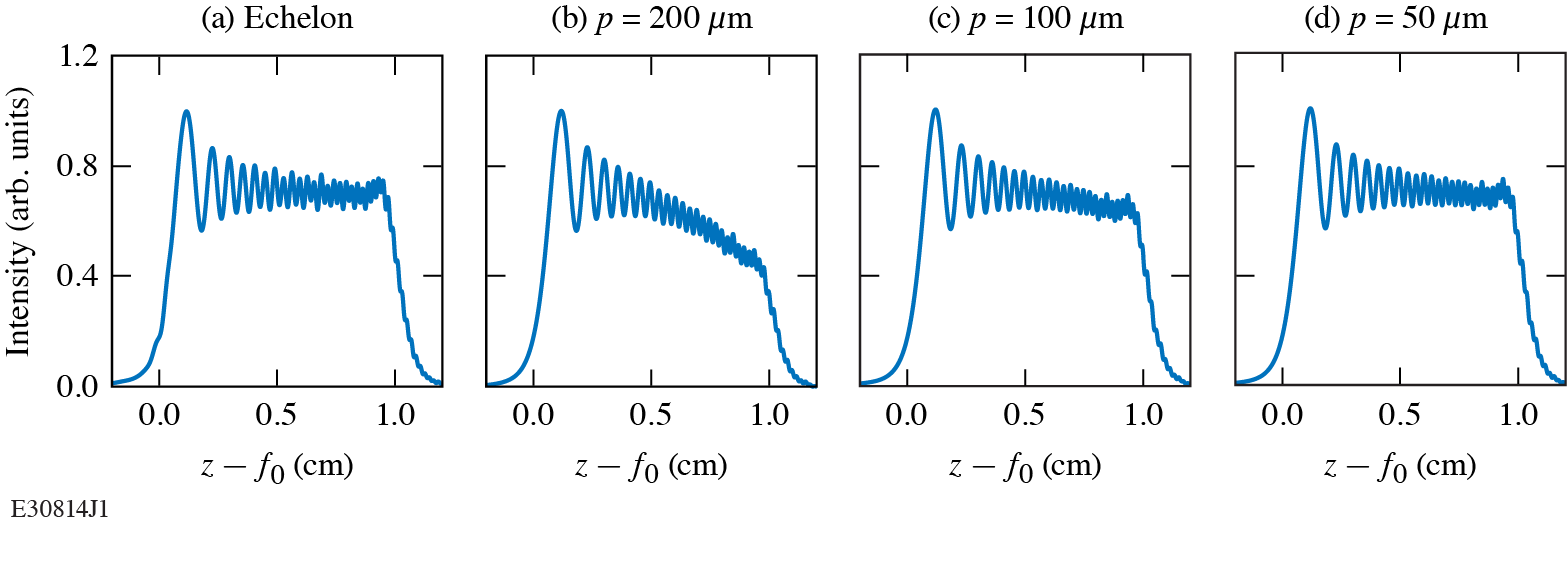}
\caption{The maximum on-axis intensity of flying focus pulses with $\varv_f = c$ created using (a) an echelon or a DM-SLM pair with an SLM pixel size of (b) $p = 200$ $\mu$m, (c) $p = 100$ $\mu$m, and (d) $p = 50$ $\mu$m.}
\label{fig:f8}
\end{figure}

A DM-SLM pair with sufficiently small pixels can create a flying focus that is virtually indistinguishable from a flying focus created by an echelon [Fig. \ref{fig:f8}]. While an echelon flattens the phase fronts globally and locally, an SLM can only flatten the phase fronts globally. Within each pixel, the phase fronts remain curved [Fig. \ref{fig:f4}(d) inset]. As a result, the constraint $\partial_{r}\phi^{\text{dm-slm}}_D(\omega,r)|_{\omega=\omega_0} = 0$ is only approximately satisfied. When the SLM pixel size is too large, the local curvature of the phase fronts affects the structure of the flying focus pulse in the far field. The inequality 
$\text{max}(\partial_r \phi_D^{\text{dm-slm}})p \ll 1$ provides a rough condition for the SLM pixel size required to reproduce the flying focus created with an echelon.  Failing to meet this condition in the near field results in a decreased intensity at corresponding locations in the far field [cf. Figs. \ref{fig:f8}(b) and (c)]. As the pixel size is reduced, the intensity profile converges to the profile produced using an echelon [cf. Figs. \ref{fig:f8}(a) and (d)]. 

\section{Examples of ultrashort flying focus trajectories} \label{profiles}
This section presents examples that demonstrate the flexibility and far-field properties of the ultrafast flying focus. The examples, i.e., constant-velocity, accelerating, and oscillating focal trajectories, are motivated by applications in plasma physics and nonlinear optics. The propagation of pulses that exhibit these trajectories was simulated in the near and far fields using a combination of the Fresnel diffraction integral and the modified paraxial wave equation (see Appendix B for details) \cite{Zhu2012,PalastroIWAVs2018}. In all cases, an axiparabola with \textit{$f_0=$} 50 cm, $R=$ 5 cm, and $L=$ 1 cm, a deformable mirror with a 5 cm radius, and a spatial light modulator with a pixel size of $p =$ 50 $\mu$m were used to extend the focal range and structure the RGD. The parameters were chosen based on the capabilities of current technology. 

\subsection{Constant-velocity focal trajectories} \label{constant profiles}

\begin{figure}[H]
\centering\includegraphics[width=\textwidth]{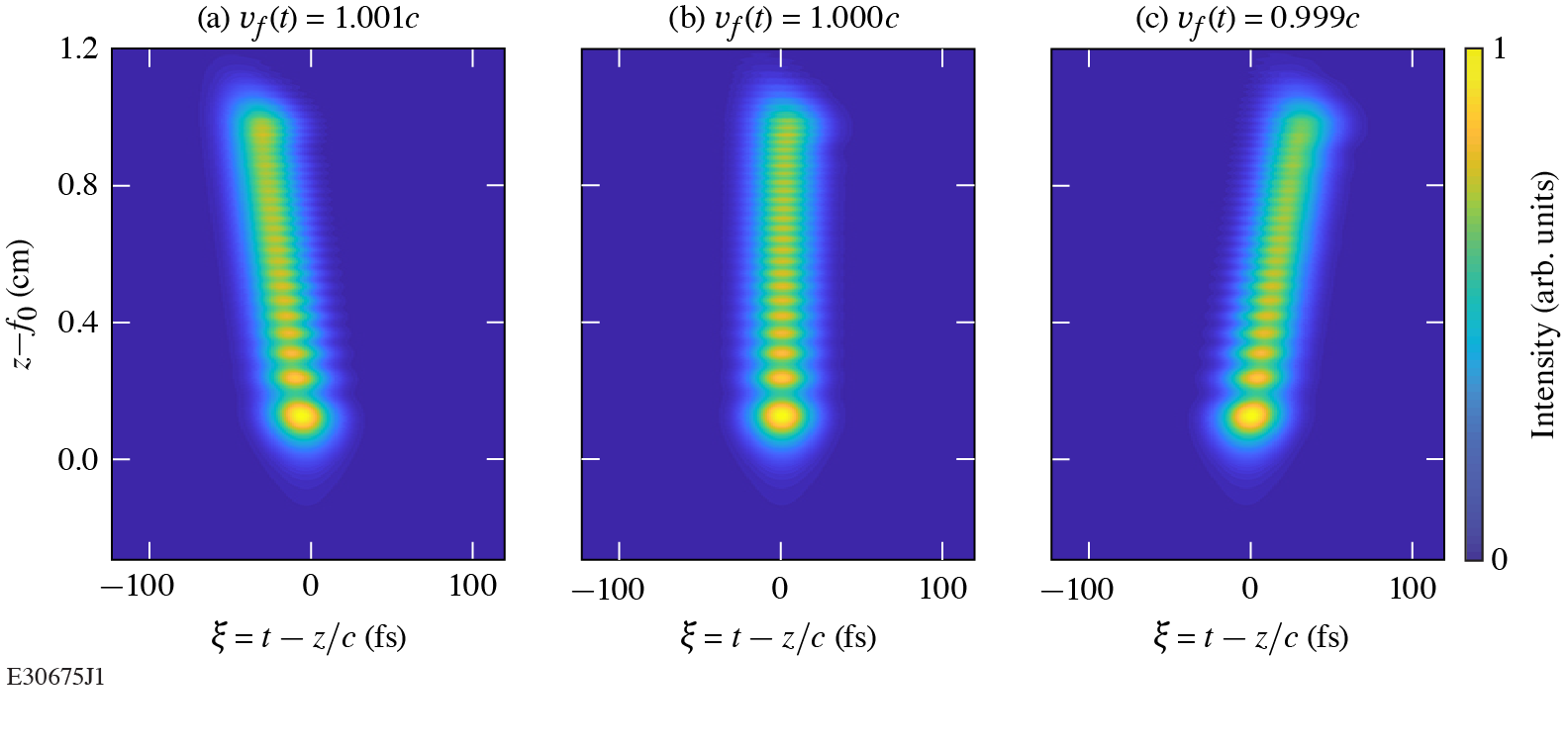}
\caption{Ultrafast flying foci with constant velocities. The maximum on-axis intensity of the pulse as a function of distance from the nominal focal point $z=f_0$ for (a) $\varv_f = 1.001c$, (b) $\varv_f = c$, and (c) $\varv_f = 0.999c$.}
\label{fig:f5}
\end{figure}

A constant-velocity flying focus can enhance applications that rely on velocity matching over long distances, such as laser wakefield acceleration \cite{Palastro2020,Caizergues2020,Debus2019,Geng2022}, THz generation \cite{Simpson2023}, and photon acceleration \cite{Howard2019,Franke2021}. Figure \ref{fig:f5} shows the on-axis intensity for the (a) superluminal, (b) luminal, and (c) subluminal velocities described in Fig. \ref{fig:f3}. In each case, the intensity peak travels along the designed constant-velocity trajectory. The images also reveal that the combination of the DM-SLM and axiparabola produce features similar to those of the axiparabola alone. Namely, the on-axis intensity is modulated, and the ultrashort pulse duration is preserved over the entire focal region [cf. Fig. \ref{fig:f2}]. 

\subsection{Exotic focal trajectories}
An accelerating focus can be used to control the trapping and acceleration of electrons in a laser wakefield accelerator. Initializing the intensity peak, and therefore the wakefield, with a subluminal velocity would facilitate the trapping of background plasma electrons in the plasma wave \cite{Schroeder2006,Palastro2020}.  After sufficient trapping has occurred, the intensity peak can be accelerated to a luminal or superluminal velocity. This change in velocity has the dual benefit of preventing electrons from outrunning the accelerating phase of the wakefield, i.e., dephasing, and of improving the quality of the electron bunch by eliminating unwanted trapping \cite{Palastro2021}.

\begin{figure}[ht!]
\centering\includegraphics[width=\textwidth]{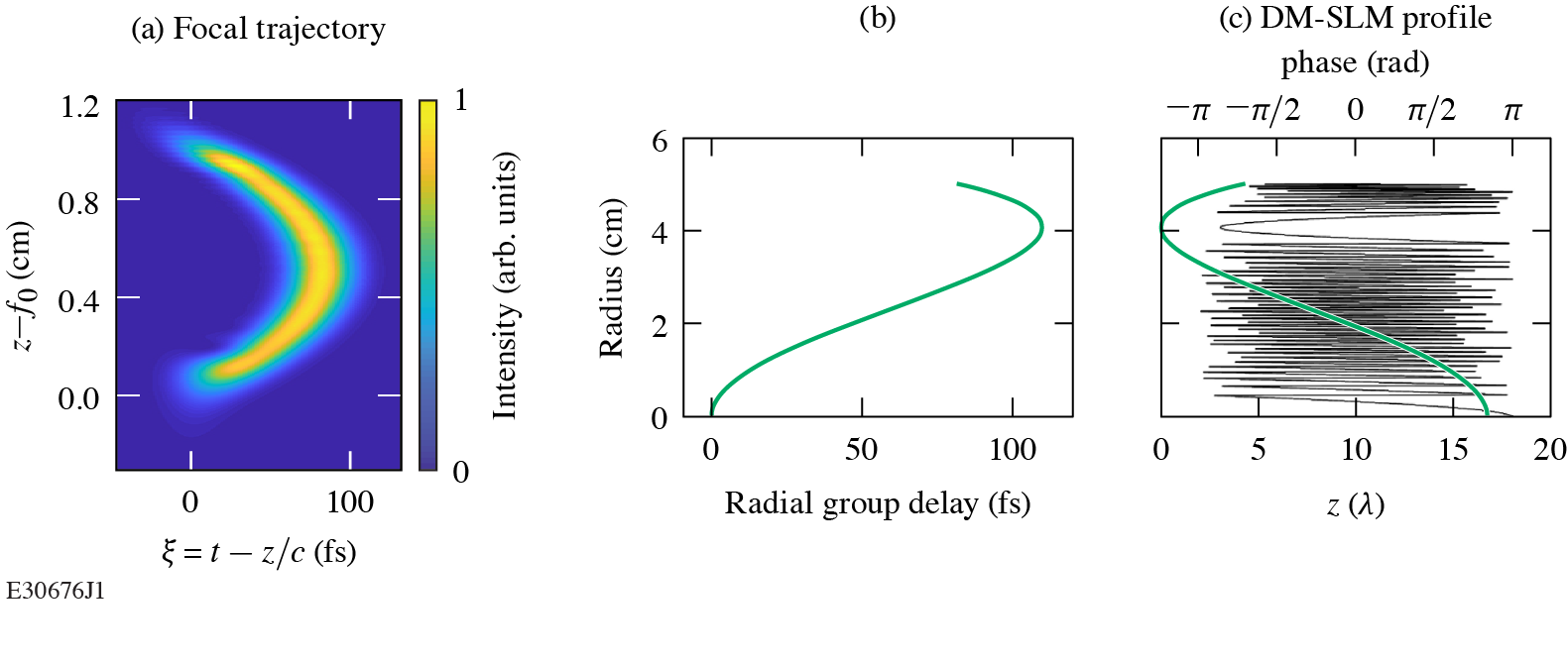}
\caption{An ultrafast flying focus that accelerates from an initial subluminal velocity to a superluminal velocity over the focal range. (a) The maximum on-axis intensity of the pulse as a function of distance from the nominal focal point $z=f_0$. (b) The radial group delay, i.e., the $\tau_D(r)$, required to produce this trajectory. (c) The corresponding deformable mirror sag function (green) and spatial light modulator phase (black). The pulse propagates from right to left.}
\label{fig:f6}
\end{figure}

Figure \ref{fig:f6} illustrates an ultrafast flying focus that accelerates from an initial subluminal velocity to a superluminal velocity over the focal range.  The design trajectory was specified as
\begin{equation} \label{eq:XLR8}
\varv_f(t) = \varv_0 + \Delta\varv\left(\frac{ct - f_0}{L}\right),
\end{equation}
with an initial velocity $\varv_0 = 0.99c$ and a velocity increment $\Delta\varv = 0.02c$.  
Over the first half of the focal range, the on-axis intensity falls back in a frame moving at the vacuum speed of light [Fig. \ref{fig:f6}(a)]. At the half-way point the velocity has increased to $c$, and thereafter the intensity peak advances in the speed of light frame. Interestingly, the radial group delay required for this trajectory [Figs. \ref{fig:f6}(b) and (c)] smooths the intensity modulations that were observed with both the axiparabola alone and with the DM-SLM constant-velocity trajectories [cf. Figs. \ref{fig:f2} and \ref{fig:f5}]. 

\begin{figure}[ht!]
\centering\includegraphics[width=\textwidth]{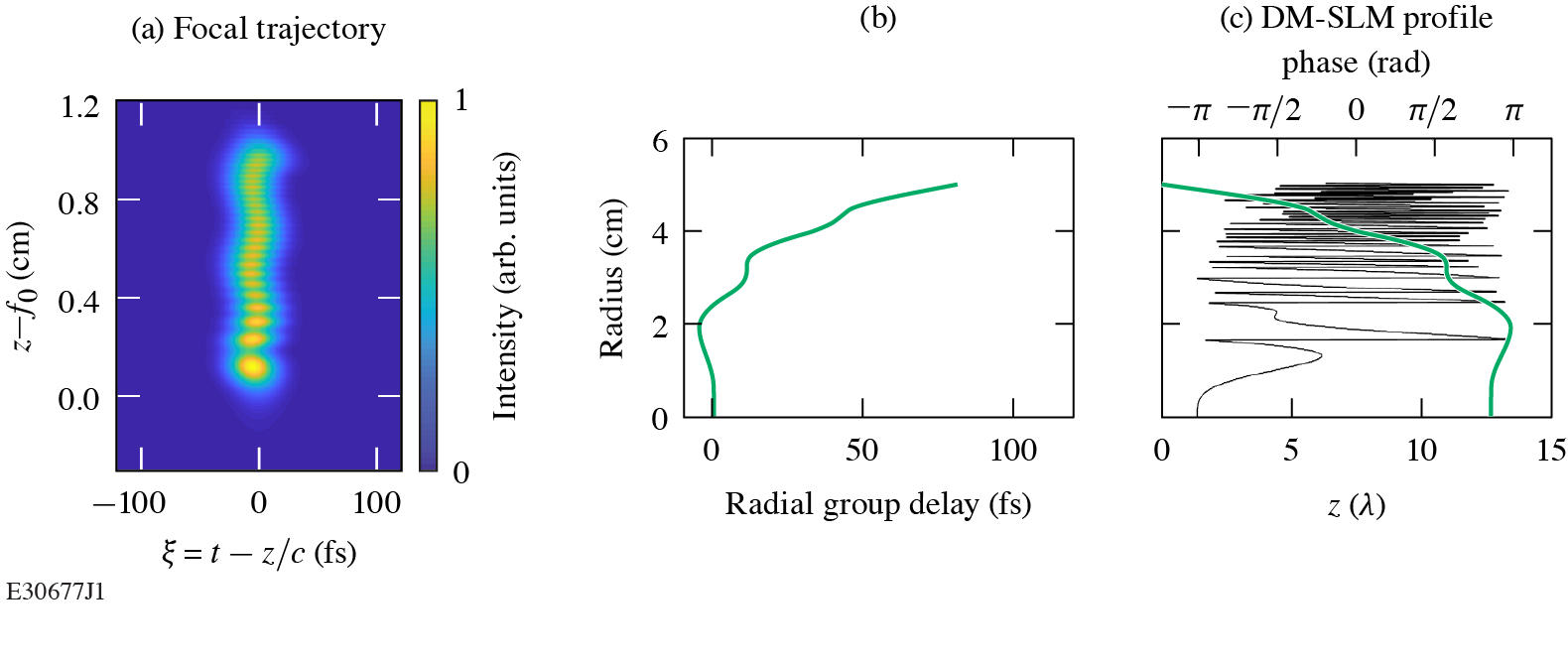}
\caption{An ultrafast flying focus that oscillates between subluminal and superluminal velocities. (a) The maximum on-axis intensity of the pulse as a function of distance from the nominal focal point $z=f_0$. (b) The radial group delay, i.e., the $\tau_D(r)$, required to produce this trajectory. (c) The corresponding deformable mirror sag function (green) and spatial light modulator phase (black). The pulse propagates from right to left.}
\label{fig:f7}
\end{figure}

A pulse with an oscillating focal point could provide a novel method for quasi-phase-matching nonlinear optical processes, a wiggler for generating radiation from relativistic electrons, or an additional degree of freedom for accessing new parametric resonances in direct laser acceleration \cite{Li2021}. An example of such a focus is shown in Fig. \ref{fig:f7}. In this case, the design focal trajectory was specified as 
\begin{equation} \label{eq:Osc}
\varv_f(t) = \varv_0 + \Delta\varv\sin\left(\frac{2\pi N(ct - f_0)}{L}\right),
\end{equation}
with a nominal velocity $\varv_0 = c$, an oscillation magnitude $\Delta\varv = 0.002c$, and $N=3$ periods. As shown in Fig. \ref{fig:f7}(a), the on-axis intensity peak oscillates between the expected velocities. While the pulse maintains its ultrashort duration, the maximum value of the intensity exhibits modulations, as it did in the case of the axiparabola alone. In general, the oscillation period of the velocity should be much greater than the Rayleigh range of the full-aperture focal spot, so that the intensity modulations do not obscure the velocity oscillations, i.e., $N\ll\pi R^2L/\lambda_0f_0^2$.

\section{Conclusions and outlook} \label{conclusion}
This work has described a method for structuring ultrashort laser pulses with dynamic focal points. The moving focal point, or ``flying focus,'' can follow a near-arbitrary trajectory over distances much greater than a Rayleigh range, while maintaining an ultrashort duration. The method employs separate optics to extend the focal range and structure the radial group delay (RGD). This overcomes a disadvantage of previous flying focus techniques, which place a lower bound on the duration of the moving intensity peak. Two specific optical configurations were considered: an axiparabola, which uses geometric aberration to extend the focal range, combined with either an echelon or a deformable mirror-spatial light modulator (DM-SLM) pair to structure the RGD. While an echelon can apply the exact RGD required for a particular focal trajectory, it is a static optic that cannot be modified on a shot-to-shot basis. The DM-SLM pair, on the other hand, has constraints imposed by the resolution of the SLM, but allows for dynamic programmability and optimization of the focal trajectory. This capability could enable rapid exploration of exotic flying foci that benefit laser-based applications in plasma physics and nonlinear optics. 

\appendix
\section{Focal trajectory produced by an extended focal range optic} \label{appendix1}
Consider a laser pulse with an initially flat phase front and flat pulse front 
propagating in the negative $\hat{\mathbf{z}}$-direction. Assuming cylindrical symmetry, the rays composing the phase and pulse front can be identified by their radial distance $r = (x^2+y^2)^{1/2}$ from the propagation axis and their frequency $\omega$. The rays travel parallel to the axis and are incident on a reflective optic defined by the sag function $s_f(r)$. At the point of reflection, each ray acquires a transverse wavenumber $k_r(\omega,r)=(\omega/c)\sin[2\theta(r)]$, where $\theta(r) = \arccos[\hat{\mathbf{z}}\cdot\hat{\mathbf{n}}(r)]$ defines the angle between the $+\hat{\mathbf{z}}$-direction and the normal vector to the surface of the optic $\hat{\mathbf{n}}(r) =  [D(r)\hat{\textbf{r}} - \hat{\textbf{z}}]/\sqrt{1+D^2(r)}$ with $D(r) \equiv ds_f/dr$. After some algebra, one finds 
\begin{equation} \label{eq:kr}
k_r(\omega,r)= - \frac{2\omega}{c} \frac{D(r)}{1+D^2(r)}.
\end{equation}
The perpendicular wavenumber is simply the radial derivative of the phase, such that
\begin{equation} \label{eq:phinonparax}
\phi_f(\omega,r) = -\frac{2\omega}{c} \int \frac{D(r)}{1+D^2(r)} dr.
\end{equation}
In the paraxial approximation, Eq. \eqref{eq:phinonparax} simplifies to $\phi_f(\omega,r)=-2\omega s_f(r)/c$, which is the first term on the right-hand side of Eq. \eqref{eq:phi}. 

The trajectory of the rays as they travel to the far field can be found by integrating the ray equations $\dot{\mathbf{x}}' = c^2\mathbf{k}/\omega$, where the overdot denotes a total time derivative and the prime denotes the instantaneous location of the ray. The radial and longitudinal locations of the rays evolve according to
\begin{align} 
r'(t) &= r + \frac{ck_r(\omega,r)}{\omega}[ct + s_f(r)] \label{eq:r'} \\
z'(t) &= s_f(r) + \frac{ck_z(\omega,r)}{\omega}[ct + s_f(r)] \label{eq:z'}, 
\end{align}
where $ct \geq -s_f(r)$, $t = 0$ corresponds to the time at which the ray with $r=0$ reflects from the optic, and $k_z(\omega,r) = [\omega^2/c^2 - k_z(\omega,r)]^{1/2}$. The focal time $t_f(r)$ and location $f(r)$ of each ray are defined as the values of $t$ and $z'$ where $r' = 0$. Solving for the value of $t$ where Eq. \eqref{eq:r'} equals zero and using this in Eq. \eqref{eq:z'} yields
\begin{align} 
ct_f(r) &= -s_f(r) + \frac{1+D^2(r)}{2D(r)}r \label{eq:tfnew} \\
f(r) &= s_f(r) + \frac{1-D^2(r)}{2D(r)}r,  \label{eq:fnew} 
\end{align}
where Eq. \eqref{eq:kr} has been used. The focal time and location are both independent of frequency.

The focal location depends implicitly on the focal time through their shared dependence on $r$. This dependence results in a focal point that moves in time. The velocity of the focal point $\tilde{\varv}_f(r)$ is given by
\begin{equation} \label{eq:vfnoparax}
\frac{\tilde{\varv}_f(r)}{c} = \frac{df}{dr}\left(\frac{dct_f}{dr}\right)^{-1} = \frac{1+D^2(r)}{1-D^2(r)},
\end{equation}
which is constrained by the focal geometry $D(r)$ and is always superluminal ($D^2$ is positive definite). 

When each ray is delayed by a time $\tau_D(r)$ before reflecting from the optic, the focal time $t_f(r) \rightarrow t_f(r) + \tau_D(r)$, and Eq. \eqref{eq:vfnoparax} can be rewritten as a differential equation for the delay needed to produce a specified focal trajectory $\varv_f(t)$:
\begin{equation} \label{eq:taunoparax}
\frac{d\tau_D}{dr} = \left[\frac{c}{\varv_f\big(t_f(r)\big)} - \left(\frac{1-D^2(r)}{1+D^2(r)}\right)\right]\frac{df}{dr},
\end{equation}
where $\varv_f\big(t_f(r)\big) = \tilde{\varv}_f(r)$. The paraxial limits of these equations are presented in the main text for simplicity.

\section{Simulation details} \label{appendix2}
The evolution of the flying focus pulse was simulated in two steps. The first step used the frequency-domain Fresnel integral to propagate the laser pulse from the flying focus optical configuration to the far field. The second step used the modified paraxial wave equation to propagate the pulse through the far field \cite{Zhu2012,PalastroIWAVs2018}. The results shown in the figures were obtained from this second step.

To solve for the evolution of the flying focus pulse, the transverse electric field was written as a carrier modulating an envelope: $\text{E}(\xi,r,z) = \frac{1}{2}e^{-i\omega_0\xi}E(\xi,r,z) + \text{c.c.}$, where $\xi = t - z/c$ is the moving frame coordinate. The carrier frequency $\omega_0$ was chosen so that the central wavelength $\lambda_0 = 2\pi c/\omega_0$ = 920 nm. The envelope $E$ was initialized just before the optical configuration in the frequency domain with the profile
\begin{equation} \label{eq:Einit}
\tilde{E}_0(\delta\omega,r) = \tilde{E}_i\Theta(r-R)\exp{(-\tfrac{1}{4}\tau^2\delta\omega^2)},
\end{equation}
where $\sim$ denotes a frequency domain field, $\delta\omega = \omega - \omega_0$, $\Theta$ is the Heaviside function, $\tilde{E}_i$ is the initial amplitude, $R$ = 5 cm, and $\tau$ = 23 fs, corresponding to a full width at half maximum duration and bandwidth of 27 fs and $\Delta \lambda$ = 78 nm, respectively. 

The phase imparted by the optical configuration, i.e., an axiparabola combined with either an echelon or a deformable mirror-spatial light modulator pair, was applied to the initial envelope. Just after the optical configuration at $z=0$, the envelope can be expressed as $\tilde{E}_0(\delta\omega,r)e^{i\phi(\omega,r)}$, where $\phi(\omega,r)$ is the phase applied by the optical configuration [Eq. \eqref{eq:phi}]. The envelope was propagated in vacuum from $z=0$ to the far-field location $z=z_i$ using the frequency-domain Fresnel integral:
\begin{equation} \label{eq:FI}
    \tilde{E}(\delta\omega,r,z=z_i) = 
    \frac{\omega}{ic z_i}\int   J_0\bigg(\frac{\omega r r'}{cz_i}\bigg)
    \exp{\bigg[\frac{i\omega(r^2+r'^2)}{2cz_i}+i\phi(\omega,r')\bigg]}\tilde{E}_0(\delta\omega,r')r' dr',
\end{equation}
where $J_0$ is the zeroth-order Bessel function of the first kind. The electric field from the Fresnel integral $\tilde{E}(\omega,r,z=z_i)$ provided the initial condition for the modified paraxial wave equation \cite{Zhu2012}: 
\begin{equation} \label{eq:MPE}
[2(i\omega_0-\partial_\xi)\partial_z + c\nabla_\perp^2]
E(r,z,\xi) = 0.
\end{equation}
The mixed space-time derivative in Eq. \eqref{eq:MPE} ensures that effects such as radial group delay and angular dispersion are modelled correctly---a requirement for accurately modeling an ultrafast flying focus. Note that Eqs. \eqref{eq:FI} and \eqref{eq:MPE} are fully consistent with one another: Eq. \eqref{eq:FI} is the integral solution to Eq. \eqref{eq:MPE}. The use of the Fresnel integral decouples the radial grids in the near field and far field, reducing computational expense compared to using  Eq. \eqref{eq:MPE} over the entire domain, especially when considering smaller $f/\#$'s \cite{PalastroIWAVs2018}.

The simulation parameters were motivated by the MTW-OPAL laser system at the Laboratory for Laser Energetics \cite{Bromage2021}, where future ultrafast flying focus experiments are being planned. The longitudinal step size $\Delta z = 2.83$ $\mu$m, temporal resolution $\Delta \xi = 0.74$ fs, and radial resolution $\Delta r = 0.60$ $\mu$m, were chosen to resolve the Rayleigh range, transform-limited pulse duration, and spot size, respectively. 

\section{On-axis intensity modulation from an axiparabola} \label{appendix3}
The Fresnel diffraction integral can be used to derive an approximate expression for the far-field, on-axis intensity profile of a laser pulse focused by an axiparabola. The expression reveals that the on-axis intensity modulations result from the spherical aberration imparted by the axiparabola and provides a condition for mitigating these modulations. The derivation begins by substituting Eq. \eqref{eq:Einit} into Eq. \eqref{eq:FI} and approximating the axiparabola phase as 
\begin{equation} \label{eq:APphase}
\phi(\omega,r') =  -\frac{\omega r'^2}{2cf_0}\left(1-\frac{L}{2f_0}\frac{r'^2}{R^2}\right),
\end{equation}
which includes the parabolic and spherical contributions and is accurate to second order in $L/f_0$. Evaluating Eq. \eqref{eq:FI} on-axis, i.e., at $r=0$, provides
\begin{equation} \label{eq:FI2}
    \tilde{E}(\delta\omega,0,z) = 
    \frac{\omega}{ic z}\int^{R}_{0}   
    \exp{\bigg[\frac{i\omega r'^2}{2c}\left(\frac{1}{z} - \frac{1}{f_0} \right)+\frac{i\omega Lr'^4}{4cf_0^2R^2} \bigg]}\tilde{E}_0(\delta\omega)r' dr',
\end{equation}
where $\tilde{E}_0(\delta\omega) = \tilde{E}_i\exp{(-\tfrac{1}{4}\tau^2\delta\omega^2)}$. Upon integrating, one finds
\begin{equation} \label{eq:FI3}
    \frac{|\tilde{E}(\delta\omega,0,z)|^2}{|\tilde{E}_0(\delta\omega)|^2} \approx \frac{\pi\omega R^2}{4cL}\left| \text{erfi}\left[\bigg(\frac{i\omega R^2}{4cLf_0^2}\bigg)^{1/2}(f_0-z)\right] - \text{erfi}\left[\bigg(\frac{i\omega R^2}{4cLf_0^2}\bigg)^{1/2}(f_0+L-z)\right] \right|^2, 
\end{equation}
where erfi is the imaginary error function and $z\approx f_0$ has been assumed. Equation \eqref{eq:FI3} oscillates with a period that varies throughout the focal region. The scale length apparent in Eq. \eqref{eq:FI3} provides a rough estimate for the modulation period: $L_M \sim (4Lf_0^2\lambda_0/R^2)^{1/2}$. The modulations can be mitigated when $L \gg L_M$ or $L \gg 4\pi Z_R$, where $Z_R = \lambda_0f_0^2/\pi R^2$ is the Rayleigh range of the full-aperture focal spot. 

\begin{backmatter}
\bmsection{Funding}
U.S. Department of Energy Office of Fusion Energy Award Number DE-SC00215057, U.S. Department of Energy National Nuclear Security Administration Award Number DE-NA0003856.

\bmsection{Acknowledgments}
The authors would like to thank D. Ramsey, J. Bromage, C. Dorrer, S.-W. Bahk, C. Jeon, B. Webb, and I. Begishev for productive discussions.

This material is based upon work supported by the Department of Energy Office of Fusion Energy under Award Number DE-SC00215057 and by the Department of Energy National Nuclear Security Administration under Award Number DE-NA0003856.
This report was prepared as an account of work sponsored by an agency of the U.S. Government. Neither the U.S.
Government nor any agency thereof, nor any of their employees, makes any warranty, express or implied, or assumes any legal liability or responsibility for the accuracy, completeness, or usefulness of any information, apparatus, product, or process disclosed, or represents that its use would not infringe privately owned rights. Reference herein to any specific commercial product, process, or service by trade name, trademark, manufacturer, or otherwise does not necessarily constitute or imply its endorsement, recommendation, or favoring by the U.S. Government or any agency thereof.

\bmsection{Disclosures}
The authors declare no conflicts of interest

\bmsection{Data Availability Statement}
Data underlying the results presented in this paper are not publicly available at this time but may
be obtained from the authors upon reasonable request.

\end{backmatter}


\bibliography{sample}

\begin{thebibliography}{10}
\newcommand{\enquote}[1]{``#1''}

\bibitem{Sainte-Marie2017}
A.~Sainte-Marie, O.~Gobert, and F.~Qu\'{e}r\'{e}, \enquote{Controlling the
  velocity of ultrashort light pulses in vacuum through spatio-temporal
  couplings,} {\protect\JournalTitle{Optica}} \textbf{4}, 1298--1304 (2017).

\bibitem{Froula2018}
D.~H. Froula, D.~Turnbull, A.~S. Davies, T.~J. Kessler, D.~Haberberger, J.~P.
  Palastro, S.-W. Bahk, I.~A. Begishev, R.~Boni, S.~Bucht, J.~Katz, and J.~L.
  Shaw, \enquote{Spatiotemporal control of laser intensity,}
  {\protect\JournalTitle{Nature Photonics}} \textbf{12}, 262--265 (2018).

\bibitem{Palastro2020}
J.~P. Palastro, J.~L. Shaw, P.~Franke, D.~Ramsey, T.~T. Simpson, and D.~H.
  Froula, \enquote{Dephasingless laser wakefield acceleration,}
  {\protect\JournalTitle{Phys. Rev. Lett.}} \textbf{124}, 134802 (2020).

\bibitem{Jolly2020}
S.~W. Jolly, O.~Gobert, A.~Jeandet, and F.~Qu\'{e}r\'{e}, \enquote{Controlling
  the velocity of a femtosecond laser pulse using refractive lenses,}
  {\protect\JournalTitle{Opt. Express}} \textbf{28}, 4888--4897 (2020).

\bibitem{Pigeon2023}
J.~J. Pigeon, P.~Franke, M.~Lim Pac~Chong, J.~Katz, R.~Boni, C.~Dorrer, J.~P.
  Palastro, and D.~H. Froula, \enquote{Interferometric measurements of the
  focal velocity and effective pulse duration of an ultrafast ‘flying
  focus’,}  (CLEO Conference, 2023).

\bibitem{Turnbull2018}
D.~Turnbull, P.~Franke, J.~Katz, J.~P. Palastro, I.~A. Begishev, R.~Boni,
  J.~Bromage, A.~L. Milder, J.~L. Shaw, and D.~H. Froula, \enquote{Ionization
  waves of arbitrary velocity,} {\protect\JournalTitle{Phys. Rev. Lett.}}
  \textbf{120}, 225001 (2018).

\bibitem{Franke2019}
P.~Franke, D.~Turnbull, J.~Katz, J.~P. Palastro, I.~A. Begishev, J.~Bromage,
  J.~L. Shaw, R.~Boni, and D.~H. Froula, \enquote{Measurement and control of
  large diameter ionization waves of arbitrary velocity,}
  {\protect\JournalTitle{Opt. Express}} \textbf{27}, 31978--31988 (2019).

\bibitem{Kabacinski2023}
A.~Kabacinski, E.~Oliva, F.~Tissandier, J.~Gautier, M.~Kozlová, J.-P. Goddet,
  I.~A. Andriyash, C.~Thaury, P.~Zeitoun, and S.~Sebban,
  \enquote{Spatio-temporal couplings for controlling group velocity in
  longitudinally pumped seeded soft x-ray lasers,}
  {\protect\JournalTitle{Nature Photonics}} \textbf{17}, 354--359 (2023).

\bibitem{Caizergues2020}
C.~Caizergues, S.~Smartsev, V.~Malka, and T.~C., \enquote{Phase-locked
  laser-wakefield electron acceleration,} {\protect\JournalTitle{Nature}}
  \textbf{14}, 475--479 (2020).

\bibitem{Debus2019}
A.~Debus, R.~Pausch, A.~Huebl, K.~Steiniger, R.~Widera, T.~E. Cowan,
  U.~Schramm, and M.~Bussmann, \enquote{Circumventing the dephasing and
  depletion limits of laser-wakefield acceleration,}
  {\protect\JournalTitle{Phys. Rev. X}} \textbf{9}, 031044 (2019).

\bibitem{Ramsey2022}
D.~Ramsey, B.~Malaca, A.~Di~Piazza, M.~Formanek, P.~Franke, D.~H. Froula,
  M.~Pardal, T.~T. Simpson, J.~Vieira, K.~Weichman, and J.~P. Palastro,
  \enquote{Nonlinear thomson scattering with ponderomotive control,}
  {\protect\JournalTitle{Phys. Rev. E}} \textbf{105}, 065201 (2022).

\bibitem{Simpson2023}
T.~T. Simpson, J.~J. Pigeon, M.~V. Ambat, K.~G. Miller, D.~Ramsey, K.~Weichman,
  D.~H. Froula, and J.~P. Palastro, \enquote{Spatiotemporal control of
  two-color terahertz generation,} {\protect\JournalTitle{Phys. Rev. L,
  (submitted)}}  (2023).

\bibitem{Formanek2022}
M.~Formanek, D.~Ramsey, J.~P. Palastro, and A.~Di~Piazza, \enquote{Radiation
  reaction enhancement in flying focus pulses,} {\protect\JournalTitle{Phys.
  Rev. A}} \textbf{105}, L020203 (2022).

\bibitem{Piazza2021}
A.~Di~Piazza, \enquote{Unveiling the transverse formation length of nonlinear
  compton scattering,} {\protect\JournalTitle{Phys. Rev. A}} \textbf{103},
  012215 (2021).

\bibitem{PalastroIWAVs2018}
J.~P. Palastro, D.~Turnbull, S.-W. Bahk, R.~K. Follett, J.~L. Shaw,
  D.~Haberberger, J.~Bromage, and D.~H. Froula, \enquote{Ionization waves of
  arbitrary velocity driven by a flying focus,} {\protect\JournalTitle{Phys.
  Rev. A}} \textbf{97}, 033835 (2018).

\bibitem{Smartsev2019}
S.~Smartsev, C.~Caizergues, K.~Oubrerie, J.~Gautier, J.-P. Goddet, A.~Tafzi,
  K.~T. Phuoc, V.~Malka, and C.~Thaury, \enquote{Axiparabola: a
  long-focal-depth, high-resolution mirror for broadband high-intensity
  lasers,} {\protect\JournalTitle{Opt. Lett.}} \textbf{44}, 3414--3417 (2019).

\bibitem{Oubrerie2022}
K.~Oubrerie, I.~A. Andriyash, R.~Lahaye, S.~Smartsev, V.~Malka, and C.~Thaury,
  \enquote{Axiparabola: a new tool for high-intensity optics,}
  {\protect\JournalTitle{Journal of Optics}} \textbf{24}, 045503 (2022).

\bibitem{Fan2023}
Q.~Fan, S.~Wang, Y.~Chen, W.~Fan, D.~Liu, Z.~Yang, Y.~Wu, W.~Zhou, L.~Cao, and
  L.~Wei, \enquote{Design of an off-axis axiparabola with inclined wavefront
  correction to obtain a straight focal line,} {\protect\JournalTitle{Opt.
  Express}} \textbf{31}, 19266--19277 (2023).

\bibitem{Sun2015}
B.~Sun, P.~S. Salter, and M.~J. Booth, \enquote{Pulse front adaptive optics: a
  new method for control of ultrashort laser pulses,}
  {\protect\JournalTitle{Opt. Express}} \textbf{23}, 19348--19357 (2015).

\bibitem{Li&KawanakaDMSLM2020}
Z.~Li and J.~Kawanaka, \enquote{Optical wave-packet with nearly-programmable
  group velocities,} {\protect\JournalTitle{Communications Physics}}
  \textbf{3}, 2399--3650 (2020).

\bibitem{Bor1988}
Z.~Bor, \enquote{Distortion of femtosecond laser pulses in lenses and lens
  systems,} {\protect\JournalTitle{Journal of Modern Optics}} \textbf{35},
  1907--1918 (1988).

\bibitem{Nemoto1997}
K.~Nemoto, T.~Nayuki, T.~Fujii, N.~Goto, and Y.-K. Kanai, \enquote{Optimum
  control of the laser beam intensity profile with a deformable mirror,}
  {\protect\JournalTitle{Appl. Opt.}} \textbf{36}, 7689--7695 (1997).

\bibitem{Liu2013}
W.~Liu, L.~Dong, P.~Yang, X.~Lei, H.~Yan, and B.~Xu, \enquote{A zernike mode
  decomposition decoupling control algorithm for dual deformable mirrors
  adaptive optics system,} {\protect\JournalTitle{Opt. Express}} \textbf{21},
  23885--23895 (2013).

\bibitem{Weiner2000}
A.~M. Weiner, \enquote{Femtosecond pulse shaping using spatial light
  modulators,} {\protect\JournalTitle{Rev. Sci. Instrum}} \textbf{71},
  1929--1960 (2020).

\bibitem{Bahk2014}
S.-W. Bahk, I.~Begishev, and J.~Zuegel, \enquote{Precompensation of gain
  nonuniformity in a nd:glass amplifier using a programmable beam-shaping
  system,} {\protect\JournalTitle{Optics Communications}} \textbf{333}, 45--52
  (2014).

\bibitem{Zhu2012}
W.~Zhu, J.~P. Palastro, and T.~M. Antonsen, \enquote{{Studies of spectral
  modification and limitations of the modified paraxial equation in laser
  wakefield simulations},} {\protect\JournalTitle{Physics of Plasmas}}
  \textbf{19} (2012). 033105.

\bibitem{Geng2022}
P.-F. Geng, M.~Chen, X.-Z. Zhu, W.-Y. Liu, Z.-M. Sheng, and J.~Zhang,
  \enquote{{Propagation of axiparabola-focused laser pulses in uniform
  plasmas},} {\protect\JournalTitle{Physics of Plasmas}} \textbf{29} (2022).
  112301.

\bibitem{Howard2019}
A.~J. Howard, D.~Turnbull, A.~S. Davies, P.~Franke, D.~H. Froula, and J.~P.
  Palastro, \enquote{Photon acceleration in a flying focus,}
  {\protect\JournalTitle{Phys. Rev. Lett.}} \textbf{123}, 124801 (2019).

\bibitem{Franke2021}
P.~Franke, D.~Ramsey, T.~T. Simpson, D.~Turnbull, D.~H. Froula, and J.~P.
  Palastro, \enquote{Optical shock-enhanced self-photon acceleration,}
  {\protect\JournalTitle{Phys. Rev. A}} \textbf{104}, 043520 (2021).

\bibitem{Schroeder2006}
C.~B. Schroeder, E.~Esarey, B.~A. Shadwick, and W.~P. Leemans,
  \enquote{{Trapping, dark current, and wave breaking in nonlinear plasma
  waves},} {\protect\JournalTitle{Physics of Plasmas}} \textbf{13} (2006).
  033103.

\bibitem{Palastro2021}
J.~P. Palastro, B.~Malaca, J.~Vieira, D.~Ramsey, T.~T. Simpson, P.~Franke,
  J.~L. Shaw, and D.~H. Froula, \enquote{{Laser-plasma acceleration beyond wave
  breaking},} {\protect\JournalTitle{Physics of Plasmas}} \textbf{28} (2021).
  013109.

\bibitem{Li2021}
F.-Y. Li, P.~K. Singh, S.~Palaniyappan, and C.-K. Huang, \enquote{Particle
  resonances and trapping of direct laser acceleration in a laser-plasma
  channel,} {\protect\JournalTitle{Phys. Rev. Accel. Beams}} \textbf{24},
  041301 (2021).

\bibitem{Bromage2021}
J.~Bromage, S.-W. Bahk, M.~Bedzyk, I.~A. Begishev, S.~Bucht, C.~Dorrer,
  C.~Feng, C.~Jeon, C.~Mileham, R.~G. Roides, K.~Shaughnessy, M.~J. Shoup~III,
  M.~Spilatro, B.~Webb, D.~Weiner, and J.~D. Zuegel, \enquote{Mtw-opal: a
  technology development platform for ultra-intense optical parametric
  chirped-pulse amplification systems,} {\protect\JournalTitle{High Power Laser
  Science and Engineering}} \textbf{9}, e63 (2021).

\end{thebibliography}

\end{document}